# 4×2 Hot electron bolometer mixer arrays for detection at 1.46, 1.9 and 4.7 THz for a balloon borne terahertz observatory

José R. G. Silva, Wouter M. Laauwen, Behnam Mirzaei, Nathan Vercruyssen, Matvey Finkel, Menno Westerveld, Nikhil More, Vitor Silva, Abram Young, Craig Kulesa, Christopher Walker, Floris van der Tak and Jian Rong Gao

*Abstract*— We have demonstrated three 4×2 hot electron bolometer (HEB) mixer arrays for operation at local oscillator (LO) frequencies of 1.46, 1.9 and 4.7 THz, respectively. They consist of spiral antenna coupled NbN HEB mixers combined with elliptical lenses. These are to date the highest pixel count arrays using a quasi-optical coupling scheme at supra-THz frequencies. At 1.4 THz, we measured an average double sideband mixer noise temperature of 330 K, a mixer conversion loss of 5.7 dB, and an optimum LO power of 210 nW. The array at 1.9 THz has an average mixer noise temperature of 420K, a conversion loss of 6.9 dB, and an optimum LO power of 190 nW. For the array at 4.7 THz, we obtained an average mixer noise temperature of 700 K, a conversion loss of 9.7 dB, and an optimum LO power of 240 nW. We found the arrays to be uniform regarding the mixer noise temperature with a standard deviation of 3-4%, the conversion loss with a standard deviation of 7-10%, and optimum LO power with a standard deviation of 5-6%. The noise bandwidth was also measured, being 3.5 GHz for the three arrays. These performances are comparable to previously reported values in the literature for single pixels and also other detector arrays. Our arrays meet the requirements of the Galactic/Extra-Galactic ULDB Spectroscopic Terahertz Observatory (GUSTO), a NASA balloon borne observatory, and are therefore scheduled to fly as part of the payload, which is expected to be launched in December 2023.

## I. INTRODUCTION

The supra-terahertz (THz) frequency range between 1 and 6 THz is very interesting and important for astronomy because it is rich in diagnostic atomic fine structure lines (e.g., [CII], [NII], [OI]), high-J lines of heavy molecules (e.g., CO) and ground-state lines of hydrides (e.g., H2O, HD) [1]. With the use of high resolution spectroscopic techniques based on a heterodyne receiver, it is possible to measure not only the line intensity but also resolve the frequency line profile that allows to extract information regarding the velocities of interstellar gas clouds. With such detailed information one can unveil the dynamics and processes that dominate, for example, in regions of star and planet formation [2-5]. Another application of heterodyne receivers is interferometry, required to spatially resolve the objects with a high angular resolution of sub arcsec [6].

The core of a heterodyne receiver is a mixing element, where the sky signal is mixed with a strong and well-known signal from a local oscillator (LO). During the mixing, the sky signal is down converted to the frequency difference between the sky and LO signals, named intermediate frequency (IF). This IF signal is in the GHz range, which makes it easier to be amplified and processed by common electronics. Furthermore, a sufficiently large IF bandwidth is needed for the detection of at least an entire spectral line without the need to tune the LO frequency. At THz frequencies the best performing mixer devices are the superconductor-isolator-superconductor (SIS) junctions [7-8] and the hot electron bolometers (HEBs) [9-10]. SIS mixers are based on photon-assisted tunnelling in the junction, having the highest sensitivity and largest IF bandwidth among the two types of mixers [11]. Because of this, up to 1 THz SIS mixers are the detector of choice for heterodyne instruments, such as those for Atacama Large Millimeter/submillimeter Array (ALMA) [12]. However, above 1 THz, the performance of such a mixer degrades rapidly due to the finite energy gap of the superconductor used, combined with the increase with frequency of the parasitic reactance of the junction. HEBs are based on the bolometric effect, where a change in the temperature induces a change in the resistance. They do not suffer from the upper frequency

This paragraph of the first footnote will contain the date on which you submitted your paper for review. This work was supported in part by the National Aeronautics and Space Administration (NASA)'s GUSTO funding through the University of Arizona and EU Horizon 2020 RadioNet. (Corresponding author: Jose R. G. Silva; Jian Rong Gao).

J. R. G. Silva and F. Van Der Tak are with the SRON Netherlands Institute for Space Research, Landleven 12, 9747 AD. Groningen, The Netherlands, and also with the Kapteyn Astronomical Institute, University of Groningen, Landleven 12, 9747 AD, Groningen, The Netherlands (e-mail: j.r.g.d.silva@sron.nl; F.F.S.van.der.Tak@sron.nl)

W. M. Laauwen, M. Finkel, M. Westerweld N. More and V. Silva are with the SRON Netherlands Institute for Space Research, Landleven 12, 9747 AD. Groningen, The Netherlands (e-mail: W.M.Laauwen@sron.nl; m.finkel@gmail.com; M.J.Westerveld@sron.nl; nmore@mpe.mpg.de; v.m.batista.da.silva@sron.nl).

B. Mirzaei, N. Vercruyssen and J. R. Gao are with the SRON Netherlands Institute for Space Research, Niels Bohrweg 4, 2333 CA, Leiden, and also with Optics Research Group, Imaging Physics department, Delft University of Technology, The Netherlands (e-mail: B.Mirzaei@tudelft.nl; nathan.vercruyssen@gmail.com; j.r.gao@sron.nl)

A. Young, C. Kulesa, and C. Walker are with the Steward Observatory, 933 N Cherry Ave., Rm N204, University of Arizona, Tucson, Arizona 85721, USA (e-mail: young@physics.arizona.edu; ckulesa@arizona.edu; iras16293@gmail.com)

2limitation, in contrast to SIS, which makes them the mixer of choice for the heterodyne instruments that operate above 1 THz. The best performing HEBs are so far based on the superconducting niobium nitride (NbN) [10], and have been demonstrated up to 5.3 THz [13]. NbN HEBs on Si substrates, with reasonable receiver noise temperatures have shown a typical IF bandwidth of 3-4 GHz [14-15]. Such devices have been previously used in instruments such as HIFI on the Herschel Space Observatory [15-16], the STO-2 balloon borne observatory [17] and upGREAT on the SOFIA air borne observatory [18]. As local oscillators, different coherent source types have been employed depending on the target frequency. For frequencies below ~2 THz the preferred LOs are solid-state sources based on frequency multiplier-chains [19-20] because they can be operated at room temperature and have a sufficiently wide frequency tuning range of ≥ 15% [20]. Starting from ~2 THz, quantum cascade lasers (QCLs) [21-22] dominate because they can be operated at any frequency within the range between 1.3 to 5 THz with sufficient output power. The frequency for the QCLs that were used or are suitable as LO can be tuned electrically by varying the bias voltage. However, the tuning range is so far limited to ≤ 10 GHz [23], which is only a small fraction of the operating frequency. With novel approaches, e.g., a QCL by applying a metasurface in combination with vertical-external cavity surface-emitting-laser structure, a 20% fractional tuning is possible [24]. Furthermore, THz QCLs are typically operated at temperatures between 40-70 K, which can be provided by Stirling coolers.

With improvements recently made to the mixer technology development, the sensitivity of NbN HEBs has reached the levels that are only a few times the quantum limit (hf/2k), which is defined for the single sideband. For example, at 5.3 THz the DSB mixer noise temperature as low as 4.2 times hf/2k has been reached [13]. Further improvements to a DSB receiver noise temperature, if possible, have a limited room with maximally a factor of ~2 at the high end of the supra-THz frequency range, as suggested in [13]. On the other hand, some of the sources or structures of astronomical interest, e.g., the Interstellar medium, are extended over angular scales much larger than the field of view of a telescope, which need to be scanned or mapped. A single pixel receiver placed at the focal point of the telescope is relatively inefficient as it samples only a small part of the field of view of the telescope. In this case, a multi pixel detector array positioned at the focal plane of a telescope can therefore increase the efficiency of the observatory [25], where the mapping speed of the instrument scales roughly with the number of pixels in the array [26-27]. However, only recent advances in some critical technologies have made receiver systems using multi-pixel arrays possible for airborne [18], balloon borne [28-30], and proposed instruments concepts for future space THz observatories [31-33]. The critical technologies include frequency multiplier-chain based multi-beam LOs [34], high power QCLs [23,35], and LO multiplexing schemes based on Fourier phase gratings [36-37].

Until now only the STO-2 [17] and upGREAT [18] instruments have made use of detector arrays at supra-THz frequencies. STO-2 employed 2-pixel arrays at 1.4 and 1.9 THz, consisting of quasi-optically coupled HEB mixers, namely using a lens-antenna scheme. upGREAT used to operate a 14-pixel array at 1.9 THz, which consisted in practice of two 7-pixel arrays for detecting two orthogonal polarizations. Besides, upGREAT has also a 7-pixel array at 4.7 THz. The upGREAT mixer arrays are based on feedhorn-waveguide structures to couple the radiation from free space to the HEB and are comprised of multiple individual mixers on physically separated blocks. In other words, they are not built on a monolithic block. Such mixers have the advantage of being easier to align and match with the instrument optics, however, they occupy a relatively large volume in the instrument, which is in contrast to the need of space instruments, where small and compact arrays are preferred. Furthermore, with this approach it is hard to realize a much larger array e.g., 64 pixels. Recent work shows potential for monolithic waveguide blocks [38], but no such mixer array has been demonstrated yet.

GUSTO [28-29] is a NASA balloon borne THz observatory that aims at exploring the inner dynamics of the Milky Way and the Large Magellanic Cloud using three heterodyne array receivers to map the fine structure lines of [NII] at 1.46 THz, referred to as Band 1 (B1), [CII] at 1.9 THz (B2) and [OI] at 4.7 THz (B3). GUSTO will use compact 4x2 HEB mixer arrays. As LOs, for B1 and B2, frequency multiplier chain arrays will be used which are developed by Virginia Diodes Inc, Charlottesville in USA [34]. However, the B3 receiver will make use of a multi beam LO in a 4x2 pattern generated using a QCL developed by MIT at Cambridge in USA [23], which is multiplexed by an asymmetric Fourier phase grating developed by SRON/TUDelft [37].

The focus of this paper is the characterization of 4x2 HEB mixer arrays to be used for GUSTO. We focus our characterization on the DSB mixer noise temperature ($T_{mixer}^{DSB}$), the mixer conversion loss ($L_{mixer}^{DSB}$) and the optimum LO power requirement ($P_{LO}$) at the mixer array level. These parameters represent the figures of merit used in the requirements set for the HEB mixers needed for GUSTO. The goal is to meet the instrument performance requirements. These arrays use a quasi-optical coupling scheme based on an elliptical lens combined with a logarithmic spiral antenna, making them the largest quasi-optical mixer arrays in the supra-THz region. The architecture used in our arrays also enables to seamlessly scale into high pixel count (>64 pixels) as will be discussed at the end of the paper.

Our paper is organized as follows. In Section II we start by introducing the instrument requirements, the different array architectures and then describe the assembled detector arrays. In Section III we highlight the experimental setup used to characterize the arrays. Section IV presents the characterization results of the arrays. The paper ends with the conclusions.

TABLE 1
GUSTO RF AND IF REQUIREMENTS FOR EACH PIXEL IN THE HEB MIXER ARRAYS. $T_{mixer}^{DSB}$ AND $L_{mixer}^{DSB}$ ARE THE PIXEL NOISE TEMPERATURE AND CONVERSION LOSS, RESPECTIVELY, DEFINED IN FRONT OF THE LENS, SEE MAIN TEXT. $T_{mixer}^{DSB}$ IS DEFINED AT AN IF FREQUENCY OF 2 GHz. $P_{LO}$ IS THE MIXER OPTIMUM LO POWER AT THE HEB.

| Lens type | Operating Frequency (THz) | $T_{mixer}^{DSB}$ (K) | $L_{mixer}^{DSB}$ (dB) | $P_{LO}$ (nW) | IF bandwidth (GHz) |
|---|---|---|---|---|---|
| B1 | 1.46 | 650 | 10.5 | 155-270 | 3 |
| B2 | 1.9 | 650 | 10.5 | 155-270 | 3 |
| B3 | 4.7 | 700 | 11 | 155-270 | 4 |

## II. HEB Mixer Arrays

In Table 1 we summarize the performance requirements of the HEB mixer arrays for GUSTO regarding sensitivity, LO power and IF bandwidth. In terms of the sensitivity, the GUSTO instrument targets an average single sideband (SSB) system noise temperature ($T_{sys}^{SSB}$) of 2900 K at 1.46, 2700 K at 1.9 THz, and 3000 K at 4.7 THz. These requirements can be broken down into allocations at the mixer array level in the form of a pair, $T_{mixer}^{DSB}$ and $L_{mixer}^{DSB}$, for each array, shown in Table 1. Here we define $T_{mixer}^{DSB}$ as the noise temperature after correcting for the optical losses in front of the lens and the noise contribution from the IF chain. This value is defined at an IF frequency of 2 GHz. We focus on quantifying and discussing both $T_{mixer}^{DSB}$ and $L_{mixer}^{DSB}$ because they are intrinsic to the array, being independent of the optics used in a test setup.

All HEB mixer arrays were designed to allow for eight pixels in a 4×2 configuration within a single metal block. All the pixels in an array share the same basic configuration that is shown in Fig. 1a. For each pixel THz radiation is collected on the surface of the elliptical Si lens. It is then focused, as it propagates through the lens and HEB chip substrate, to the spiral antenna, where the radiation is converted to an AC electrical current that is fed to the HEB. Through bonding wires, the HEB is connected to a co-planar waveguide (CPW) line that is used to both DC bias the device and collect the IF signal from the mixer. Each pixel is terminated with an IF connector that acts as the interface to a low noise amplifier (LNA). In the array the IF lines are placed such that no IF cross talk is present in the assembled circuit board. To confirm this, we measured a S21 < -60dB, where S21 represents the power transferred from Port 1 to Port 2, with each port in our case being a different IF line.

The lenses and the substrate of the HEB chips are made of pure, highly resistive Si (≥5 kΩ cm), which has a negligible optical loss at cryogenic temperatures [39]. Each HEB chip consists of a NbN bridge integrated with a planar logarithmic spiral antenna. We chose such an antenna because it has a high power coupling efficiency to the HEB bridge over a wide range from 1 THz to 5.3 THz [13,40,41]. Other options, such as twin slot antennas, have never been demonstrated for low noise HEB mixers above 2.5 THz [42]. Additionally, because of the wide-band coverage of a spiral antenna, we can apply a common design for the arrays operated at the three different frequencies, which can reduce the cost significantly. The antenna structure is similar to the one used in [40]. Some details of the antenna are given in [43]. Elliptical lenses were chosen since they offer high coupling of the radiation to an antenna and also higher gaussicity of the beam compared to a hemispherical lens [44].

Two models of detector arrays were designed to accommodate the two types of lenses with different diameters. In Fig. 1b we show the completed B1 and B2 arrays, using 10 mm diameter lenses and having a pitch size of 11 mm. The two arrays make use of the same model and were optimized for operation at 1.46 and 1.9 THz, respectively. Because these two frequencies are very close, it is difficult to separate them in the optical path of the instrument. Thus, B1 and B2 were designed to be placed side by side on the cold plate of the cryostat, mimicking a 4×4 array. The devices used have a NbN bridge of 2 μm in width, 0.15 μm in length, and 5 nm in thickness.

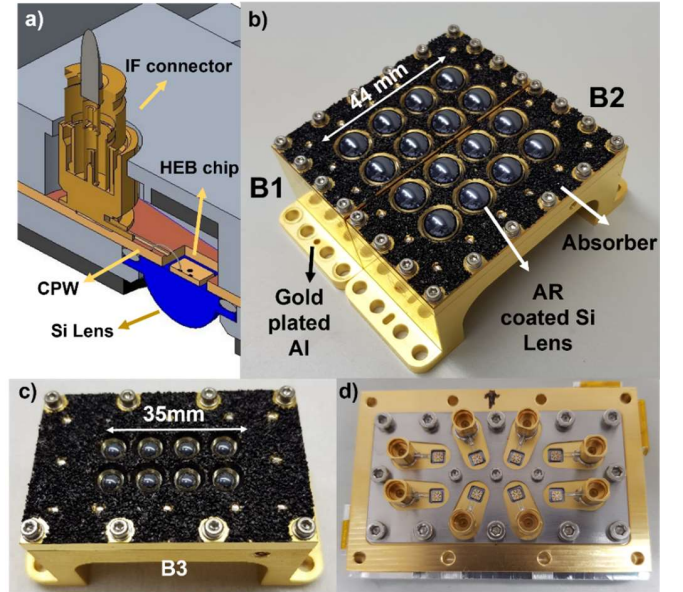

Fig. 1. 4×2 HEB mixer arrays. a) Schematic of the single pixel configuration used in all the arrays. THz radiation is collected at the elliptical surface of the lens and focused to the HEB antenna. The HEB is connected through bonding wires to a co-planar waveguide (CPW) transmission line, which is used to both bias the device and carry out the IF signal from the mixer. The other end of the CPW line is terminated with an IF connector that is the interface to a low noise amplifier. b) Completed B1 and B2 arrays, for operation at 1.46 and 1.9 THz, respectively. The arrays are presented side by side, mimicking a 4×4 array. This is the intended placement on the cold plate of the GUSTO instrument. c) Completed B3 array designed to operate at 4.7 THz. d) A back side view of the partly assembled B3 array, where the eight HEB chips, CPW lines and IF connectors are shown, from [47].

Besides a good impedance matching between the HEBs and the antennas, such dimensions of the HEBs provide an optimum LO power within the requirements described in Table 1. The critical temperature of the NbN bridges is about 10 K. In Fig. 1c we show the completed B3 array that uses 5 mm diameter lenses and has an 8 mm pitch size. This array is optimized for operation at 4.7 THz. In Fig. 1d we present a back side view of the B3 array, while partly assembled, where the eight detector chips, CPW lines and IF connectors are shown. The HEB devices used in this array are similar to the ones used in the other arrays (from the same wafer), however, the NbN bridge lengths are longer, being 0.2 μm instead. The increased HEB length increases the volume of the HEB, and thus the LO power required.

Each of the arrays has a different lens design, optimized to meet the GUSTO optical beam requirements. The optimization study and verification can be found in [45]. The detailed lens designs are shown in Table 2. Additionally, each lens is coated with Parylene C as an anti-reflection (AR) coating with the ideal thickness, designed using Equation 2 in [46]. Both realized (measured) and designed thicknesses of the Parylene C are also shown in Table 2. The differences are due to the limited accuracy in controlling the thickness during the coating process. The methodology used to mount and align HEB antenna with the lens optical axis has been described elsewhere [47].

TABLE 2
CHARACTERISTIC PARAMETERS OF THE LENSES AND AR COATING FOR THREE DIFFERENT FREQUENCIES. EACH LENS TYPE IS USED IN A DIFFERENT ARRAY. THE EXTENSION LENGTH INCLUDES THE DETECTOR SUBSTRATE, WHICH HAS A THICKNESS OF 342 ±2 μM. FOR THE PARYLENE-C WE PRESENT FIRST THE REALIZED THICKNESS AND IN PARENTHESES THE IDEAL, DESIGNED THICKNESS.

| Lens type | Operating Frequency (THz) | Major axis (μm) ±2 μm | Minor axis (μm) ±2 μm | Extension length (μm) ±2 μm | Parylene-C thickness (μm) ±0.2 μm |
|---|---|---|---|---|---|
| B1 | 1.46 | 5235 | 5000 | 1542 | 33 (31.7) |
| B2 | 1.9 | 5235 | 5000 | 1527 | 24.5 (24.4) |
| B3 | 4.7 | 2617 | 2500 | 767 | 9.0 (9.8) |

### III. EXPERIMENTAL SETUP

We measure the DSB receiver noise temperature ($T_{rec}^{DSB}$), the receiver conversion loss ($L_{rec}^{DSB}$), and the required LO power ($P_{LO}$) for each pixel in the arrays. The measurements for the three arrays were performed at 1.39, 1.63 and 5.25 THz, respectively, being slightly different from GUSTO's respective B1, B2, and B3 center frequencies. Since we do not have the same LOs as GUSTO available at SRON (where the experiments were performed), the choice of characterization frequency for the different mixer arrays was limited to the closest THz lines available from the far infrared (FIR) gas laser used as LO in our heterodyne measurement setup. The IF noise bandwidth (NBW) was measured in the IF frequency range between 0.5 and 5 GHz for a few selected mixers. Additionally, the beam properties and pointing direction of the mixers were also characterized and can be found elsewhere [45,47].

The heterodyne measurement setup used in our experiments is schematically presented in Fig. 2. The LO is a FIR gas laser operated at 1.39, 1.63 or 5.25 THz. We use a swing arm attenuator in combination with a proportional-integral-derivative feedback loop to stabilize or sweep the LO power when measuring $T_{rec}^{DSB}$ [48]. The radiation from both the LO and the blackbody load, being either hot (at a temperature of 290-295 K) or cold (77K), are combined with a 3 μm thick Mylar beam splitter. The combined radiation propagates through a 1.2 mm thick ultra-high molecular-weight polyethylene (UHMWPE) cryostat window and a QMC heat filter with a cut-off frequency of 5.8 THz at 4 K, to the lens of the pixel being measured. The total air distance between the hot or cold load and the window of the cryostat is ≈30 cm. In the schematic we also show an array, which is mounted on the 4K plate of the cryostat. However, only one pixel could be measured at a time since we could not perform the measurements of all the pixels simultaneously, limited by our setup. The physical temperature of the mixers during the measurements ranges between 4.3 and 4.5 K.

The IF chain consists of a bias-T and a cryogenic SiGe low noise amplifier (LNA) [49]. The latter is connected thermally to the 4 K plate. The room temperature part of the IF chain includes two LNAs, a bandpass filter, a 2.4 GHz low pass filter (LPF) and a microwave power meter. For $T_{rec}^{DSB}$ measurements the IF was filtered by the bandpass filter with a bandwidth of 100 MHz, centered at 2 GHz. The IF chain had a total gain of 85 dB and a noise temperature of 6.5 K at 2 GHz. For the noise bandwidth measurements, we replaced the components from the bandpass filter up to the power meter in Fig. 2 with a spectrum analyzer.

### IV. RESULTS AND DISCUSSION

#### A. Pixel Characterization

In Fig. 3 we present the characterization of an HEB mixer from the B3 array at 5.25 THz. Fig. 3a shows three measured current-voltage (IV) curves of the HEB in the unpumped state, when no LO is applied, and two pumping states around the optimum $P_{LO}$, where the $T_{rec}^{DSB}$ becomes the lowest. The $T_{rec}^{DSB}$ is obtained using the Y-factor technique, where for a given bias of voltage and current the receiver output powers of the HEB in response to the hot load ($P_{hot}$) and the cold load ($P_{cold}$) are measured. We then obtain the Y-factor by calculating the ratio between $P_{hot}$ and $P_{cold}$, where the Callen-Welton blackbody temperatures are used [50]. In Fig. 3b we present an example of a Y-factor measurement for the same pixel. In this case the lowest $T_{rec}^{DSB}$ is 2110 ± 100 K when the device is biased at a current of 32 µA and a voltage of 1 mV. For this particular bias point the $L_{rec}^{DSB}$, obtained using the U factor technique [51], is 13.6 ± 0.5dB [52]. Using the isothermal technique [53] we estimate the $P_{LO}$ for the same mixer to be between 192-199 nW. The optimal operation region in the IV, where we obtain less than 5% degradation of the $T_{rec}^{DSB}$, covers a voltage range between 0.6 - 1.0 mV and currents between 28-40 µA, which is highlighted in Fig. 3a.

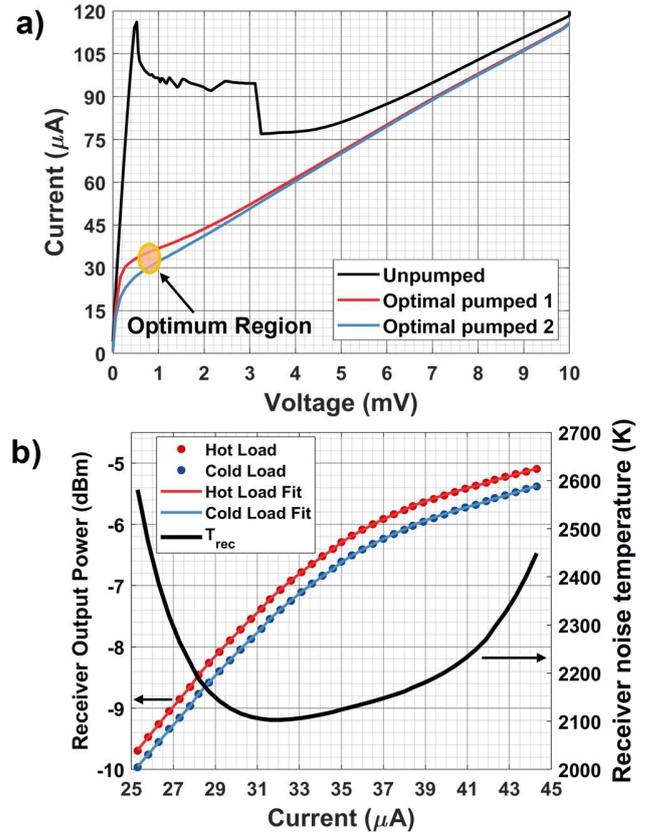

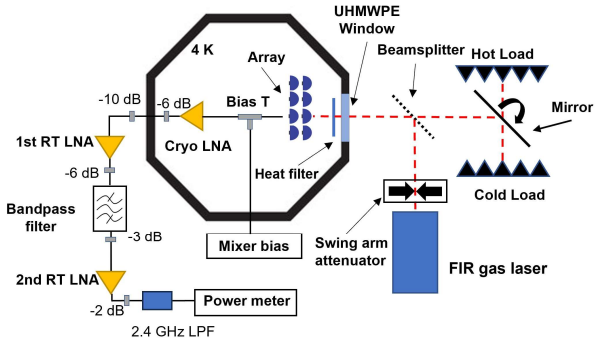

Fig. 2. Experimental setup for measuring double sideband receiver noise temperature ($T_{rec}^{DSB}$), receiver conversion loss ($L_{rec}^{DSB}$) and optimum local oscillator power ($P_{LO}$). The hot and cold loads and the beam splitter are in air. We use the rotating mirror to change between the hot and cold load. The IF noise bandwidth was measured using the same setup, but the part including the Band pass filter, 2nd RT LNA, 2.4 GHz LPF and the power meter is replaced with a spectrum analyzer.

Fig. 3. Characterization of a mixer out of the B3 array at 5.3 THz. a) Unpumped and optimally pumped current-voltage curves. Highlighted optimum region, where the $T_{rec}^{DSB}$ degrades less than 5% from the lowest value. b) Measured receiver output power for both hot and cold loads and respective polynomial fit as a function of the HEB bias current, and the resulting $T_{rec}^{DSB}$. The HEB was biased at a voltage of 1 mV.

As discussed in the previous section we are interested in $T_{mixer}^{DSB}$ and $L_{mixer}^{DSB}$. To derive $T_{mixer}^{DSB}$ from the measured $T_{rec}^{DSB}$ and $L_{rec}^{DSB}$, we apply the well-established equation in [1] as follows:

$$T_{mixer}^{DSB} = \frac{T_{rec}^{DSB} - T_{Opt} - T_{IF} \times L_{rec}^{DSB}}{L_{Opt}} \quad (1)$$

where $T_{Opt}$ and $L_{Opt}$ are the noise temperature and losses, respectively, caused by the optics in the optical path between the hot/cold load and the Si lens as shown in Fig. 2, and TIF is the noise temperature from the IF chain (6.5 K). The optical losses at different frequencies in our measurements are summarized in Table 3. By applying Eq. (1), we obtain the $T_{mixer}^{DSB}$, which is essentially the result of subtracting the noise contributions from all the optics in front of the Si lens and the IF chain. For clarity, we stress again that the derived $T_{mixer}^{DSB}$ in this way includes both the optical loss of the Si lens and power coupling loss between the antenna and the bolometer. This is different from the intrinsic mixer noise temperature that is determined by the HEB itself.

Since B2 and B3 arrays were characterized at different frequencies from the GUSTO frequencies, we need to correct for the deviation caused by the difference in the frequency in order to compare with the required performance by GUSTO. For the B2 pixels we estimate a 5% higher $T_{mixer}^{DSB}$ at 1.9 THz than what measured at 1.63 THz based on the work in [13]. For the B3 pixels, we estimate a 7.5 % lower $T_{mixer}^{DSB}$ at 4.7 THz than what was obtained at 5.25 THz. The difference was established by measuring the $T_{rec}^{DSB}$ of a similar mixer at both 4.7 THz and 5.25 THz, which is described in [54]. For B1 array, the difference in frequency between the LO used for characterization and LO of B1 is so small, no corrections are applied.

TABLE 3
OPTICAL LOSSES INCLUDING THE AIR, 3 µM THICK MYLAR BEAM SPLITTER (BS), AND WINDOW AT ROOM TEMPERATURE, AND THE HEAT FILTER AT 4K IN OUR HETERODYNE MEASUREMENT SETUP. AMONG THEM, BS LOSSES ARE SIMULATED, AIR LOSS AT 5.3 THZ WAS MEASURED, WHILE THE AIR LOSSES AT THE OTHER TWO FREQUENCIES ARE SIMULATED. THE REMAINING LOSS VALUES ARE MEASURED.

| Array | LO Frequency (THz) | Air (dB) | Mylar BS (dB) | Window (dB) | Heat Filter (dB) | Total optical losses (dB) |
|---|---|---|---|---|---|---|
| B1 | 1.39 | 0.87 | 0.07 | 0.43 | 0.66 | 2.03 |
| B2 | 1.63 | 0.64 | 0.09 | 0.38 | 1.14 | 2.25 |
| B3 | 5.25 | 0.9 | 0.63 | 1.47 | 0.56 | 3.56 |

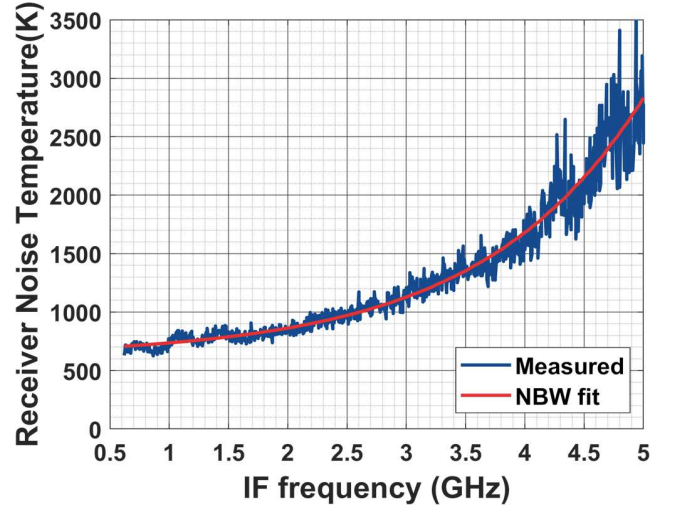

Fig. 4. NBW measurement for a B1 pixel at 1.39 THz. The measured receiver noise temperature as a function of the IF frequency was fitted with a generic exponential equation. From the fitted curve, we estimate a NBW of 3.5 GHz, defined as the frequency at which the receiver noise temperature increases by 3dB.

To illustrate how we do the corrections, we take the same B3 array pixel used for the measurements in Fig. 3, as an example. We first apply Eq. (1) to the lowest measured $T_{rec}^{DSB}$ (2100K), using $T_{opt}$ = 342K, $L_{rec}^{DSB}$ = 13.6dB and $L_{Opt}^{DSB}$ = 3.56 dB, to derive the $T_{mixer}^{DSB}$ at 5.3 THz. Afterwards by applying a reduction factor of 0.925 (corresponding to 7.5%) to the data at 5.3 THz, we derive a $T_{mixer}^{DSB}$ at 4.7 THz, which is 665 ± 40K.

To measure the NBW of an HEB mixer in our arrays we repeat the $T_{rec}^{DSB}$ measurements over a wide IF range when it is biased at an optimal operating point. In Fig. 4 we show one measurement for a B1 pixel at 1.39 THz, where the measured $T_{rec}^{DSB}$ is plotted as a function of IF frequency. By fitting a generic exponential equation, $T_{rec}^{DSB} = T_0 + a * \exp\left(\frac{f}{b}\right)$, to the measured data, we find the frequency where the fitted $T_{rec}^{DSB}$ increases by 3 dB, which determines the NBW and is 3.5 GHz. We obtained the same NBW value for a B3 pixel measured at 5.25 THz, that confirms the data at 1.39 THz. The measured NBW is sufficient to fully meet the IF bandwidth requirements for B1 and B2 arrays for GUSTO but is slightly smaller than what is required for B3. We would like to argue that this NBW is expected because it is limited by the NbN film technology, specifically due to the film thickness of 5-6 nm in practice [55,56] and a Si substrate used. The measured NBW here is close to the one previously reported in an NbN HEB produced in our labs from a different film in [57], which was 4 GHz at 4.7 THz. It also agrees with the NBW results reported for the mixers used in upGREAT, which were 4 GHz for the mixers at 1.9 and 4.7 THz [18], and those in [58], which were between 3 and 3.5 GHz.

*B. HEB Mixer arrays results*

We summarize the performance of all the pixels in the three arrays at the GUSTO operating frequencies in Fig. 5, where the $T_{mixer}^{DSB}$ of each pixel in B2 and B3 arrays has been corrected to their GUSTO operating frequencies, and where $T_{mixer}^{DSB}$ are shown in (a), $L_{mixer}^{DSB}$ in (b), and $P_{LO}$ in (c). To illustrate the array performance, we summarize the average and standard deviation value of $T_{mixer}^{DSB}$, $L_{mixer}^{DSB}$ and $P_{LO}$, for each array in Table 4.

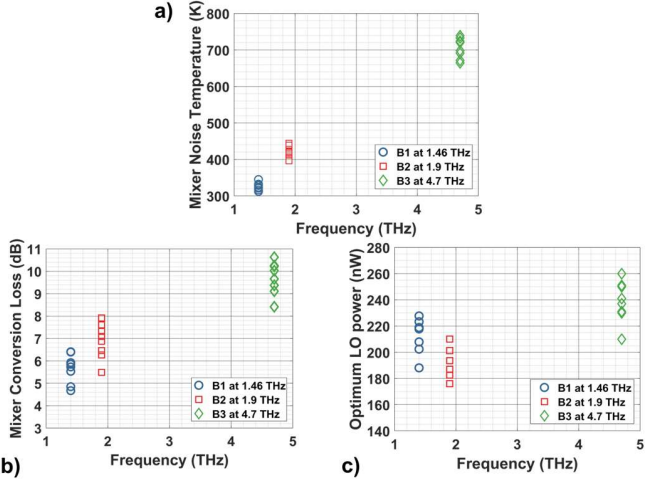

Fig. 5. Mixer noise temperature at 2 GHz IF (a), mixer conversion loss (b), and optimum LO power (c) for the different elements of three HEB mixer arrays characterized at GUSTO's frequencies, 1.46, 1.9 and 4.7 THz.

TABLE 4
PERFORMANCE SUMMARY OF THE THREE HEB MIXER ARRAYS AVERAGED OVER THE 8 PIXELS IN AN ARRAY. IT INCLUDES THE MEASURED MIXER NOISE TEMPERATURE ($T_{mixer}^{DSB}$) AND MIXER CONVERSION LOSS ($L_{mixer}^{DSB}$) AT 2 GHz IF, THE RECEIVER NOISE TEMPERATURE ($T_{rec,GUSTO}^{DSB}$), WHICH WAS ESTIMATED WHEN WE APPLY THE GUSTO OPTICS AND THE MEASURED $T_{mixer}^{DSB}$ AND $L_{mixer}^{DSB}$, AND OPTIMUM LO POWER AT HEB ($P_{LO}$). IN PARENTHESES ARE THE STANDARD DEVIATIONS WITHIN THE RESPECTIVE ARRAY.

| Array | Operating Frequency | $T_{mixer}^{DSB}$ (K) | $L_{mixer}^{DSB}$ (dB) | $T_{rec,GUSTO}^{DSB}$ (K) | $P_{LO}$ (nW) |
|---|---|---|---|---|---|
| B1 | 1.46 THz | 330 (10) | 5.7 (0.6) | ≈750 | 210 (12) |
| B2 | 1.9 THz | 420 (14) | 6.9 (0.7) | ≈1000 | 190 (10) |
| B3 | 4.7 THz | 700 (26) | 9.7 (0.7) | ≈1650 | 240 (15) |

The average $T_{mixer}^{DSB}$ for B1, B2, and B3 arrays are 330 K, 420K, and 700 K, respectively. We demonstrate the arrays uniformity on $T_{mixer}^{DSB}$ in Fig. 5a, where their standard deviations are within a range of 3 to 4% for the three arrays. An increase of the $T_{mixer}^{DSB}$ with the operating frequency is expected. However, the value at 4.7 THz is a factor of 2.1 times more than what at 1.46 THz, which is more than what reported in [13], where the factor for the two same frequencies was about 1.7. This suggests that the increase in $T_{mixer}^{DSB}$ is partly due to the contribution of quantum noise and partly due to the additional losses within the mixer, as indicated by the higher $L_{mixer}^{DSB}$ at 4.7 THz. The additional losses at 4.7 THz are expected to be caused by the loss in the antenna and the use of the smaller Si lenses (5 mm) [59]. The former will be discussed in the next paragraphs. In terms of using the unit of quantum noise (hf/2k), they are 9.4x hf/2k at 1.46 THz and 6.2x hf/2k at 4.7 THz.

For $L_{mixer}^{DSB}$, we find that it increases with the array operating frequency. Such an increase is confirmed even in the intrinsic $L_{mixer}^{DSB}$ after removing the optical loss of the lens and coupling loss between antenna and HEB. This result contradicts the expectation for NbN HEBs since it should be independent of the operating frequency as long as the LO frequency is above the gap frequency of the thin NbN [13,60]. Based on scanning electron microscopy images of some of our devices, we have noticed some artifacts that are present around the gold spiral antenna arms, which may introduce additional ohmic losses to the THz RF current. This effect would be stronger for a higher frequency and thus could introduce additional RF loss, which was not included in our analysis. Furthermore, we notice a relatively large variation on $L_{mixer}^{DSB}$ within an array in Fig. 5b, where standard deviations of 7-10% are found for the three arrays. We believe this might be caused by the absence of a circulator in the U-factor measurements, where the circulator is crucial to eliminate standing waves between the HEB and the cryogenic LNA. Without a circulator, the changes in the output impedance of an HEB can change the standing wave behavior, affecting the determination of $L_{mixer}^{DSB}$.

The $P_{LO}$ for the B3 array is slightly higher than that for B1 and B2 due to the greater length of the HEBs used in the B3 array. Within a given array the $P_{LO}$ distribution is uniform, with a standard deviation around 5-6% consistently for the three arrays. However, for B1 and B3 arrays we do have one outlier pixel for each array as shown in Fig. 5c. Nevertheless, even with these two outliers, such $P_{LO}$ uniformity is good enough, allowing to pump all the mixers in the array within their optimum operation regions, where less than 5% degradation of their $T_{mixer}^{DSB}$ is expected. A comparison of LO power required in our cases with other instruments in the literature is provided in [61].

These arrays have in principle met the performance requirements demanded by the instrument and are scheduled to fly on board of GUSTO. Here we report only three arrays, however, we have also built and characterized two backup arrays (five arrays in total). One backup array was optimized for 1.6 THz and can be used to replace either the B1 array or the B2 array, while the other one is optimized for 4.7 THz to replace the B3 array if necessary.

We now compare the mixer performance in our arrays with some of the best single pixel results reported previously in the literature, and with the performance of other instruments in the next paragraph. Our average $T_{mixer}^{DSB}$ of 330K at 2GHz IF, at 1.46 THz (for B1) is very similar to a $T_{mixer}^{DSB}$ of 300 K at 1.5 GHz IF (measured at 1.3 THz) reported by K. M. Zhou et al [62,63], which was derived from their $T_{rec}^{DSB}$ (600 K) and the optical losses. Our average $T_{mixer}^{DSB}$ of 420 K at 2 GHz IF obtained at 1.9 THz (B2) is close to what was reported in our labs, by Zhang et al [13], where a $T_{mixer}^{DSB}$ of 380 K at 1.5 GHz IF is derived. Additionally, our result for B2 array is also in line with or even better than the single pixel $T_{rec}^{DSB}$ of 900 K at 1.5 GHz IF, reported by Kloosterman et al [64], where we are not able to extract a $T_{mixer}^{DSB}$ due to missing details. Our average $T_{mixer}^{DSB}$ of 700 K at 2 GHz IF, at 4.7 THz (B3) is about 17% higher than the best value reported for a single pixel in our labs, by Kloosterman et al [65], for which we estimate a $T_{mixer}^{DSB}$ of 600 K at 2 GHz IF. The difference can be attributed to the loss in the antenna and the use of the smaller Si lenses (5 mm) as discussed previously.

Our arrays, integrated in the GUSTO instrument, have shown for B1 at 1.4 THz an averaged GUSTO receiver noise temperature, $T_{rec,GUSTO}^{DSB}$, of 870 K; for B2 at 1.9 THz an averaged $T_{rec,GUSTO}^{DSB}$ of 1100 K; and for B3 at 4.7 THz an averaged $T_{rec,GUSTO}^{DSB}$ of 1920 K. All the $T_{rec,GUSTO}^{DSB}$ values above are taken at a physical temperature of 5.1 K for the mixers and an IF frequency of 1 GHz. To compare with our expected receiver noise temperatures in table 4 the measured values should be corrected by an increase of ≈11% which represents the increase in noise temperature from 1 to 2 GHz IF (16% degradation) combined with the lower temperature of the detectors in our measurements (5% improvement). Therefore, the measured $T_{rec,GUSTO}^{DSB}$ are slightly worse than what we expected in Table 4. This can be explained by the fact that in our lab. setup for Y-factor measurements, before integration, the entire beam pattern is coupled the hot/cold load, whereas in GUSTO there is sidelobe spillover throughout the optics and especially beam vignetting in some optical elements for some of the mixers. These effects contribute to the increase of the noise temperature, and hence the differences seen. The GUSTO instrument is currently undergoing final integration with the Gondola in preparation for the launch in December 2023 and is therefore not fully ready to make a concrete comparison. During the final commissioning, the ultimate performance can be determined and a more detailed paper on GUSTO's performance will be prepared.

*C. Scaling the pixel count in HEB mixer arrays*

Although the arrays for GUSTO were designed in a 4×2 configuration, the need for both B1 and B2 arrays to be placed side by side on the cold plate of the instrument will demonstrate practically a 4×4 mixer array using our array architecture. In this case, care should be taken to ensure the pointing direction of the mixers in one array to be parallel to those from another array. The accurate pointing of 8 mixers within one array has been demonstrated [47]. Furthermore, in the case of B1 and B2 the final pointing was achieved relative to the same reference, effectively demonstrating the accurate pointing of the pixels between the two arrays.

The above approach allows to extend an array with more pixels, for example, an 8x8 pixel array. We argue that we can also build in principle eight 4x2 sub-arrays with the right mechanical adaptations, which can be assembled, characterized for their sensitivity and beam pointing independently, and then are mounted on the cold plate of an instrument, like GUSTO. In terms of the LO, for 1.46 THz, one can build eight sub-arrays of LO based on frequency multiplier chains and combine them to form 64 LO beams. At the higher frequencies, including 1.9 THz, a combination of a QCL with a phase grating could be used to generate 64 LO beams. High power QCLs have been demonstrated, for example, a 4.7 THz QCL with an output power of 8 mW at 55 K [23] and a 1.8 THz QCL with an output power of 28 mW at 10K [68]. Based on the GUSTO experience, about 10 mW will be sufficient to pump a 64 pixel array. Additionally, a phase grating to generate 81 beams from a single QCL with a high efficiency (94 %) has been demonstrated in [69], that can be applied for generating 64 beams as well. Therefore, we conclude that a large HEB array receiver of 64 pixels is feasible using current array approach and testing facilities.

To further expand the array, e.g., ≥100 pixels, one should explore an integrated approach similar to what was used for the direct detector array [70], which consists essentially of an array of lenses fabricated on a wafer and an array of HEBs also on a wafer. The mixer array is formed by aligning two wafers, that may offer a reliable and cheap technology for future large heterodyne arrays. The concept study of this new approach, including the choices of low noise amplifiers and spectrometers, has been described in detail in [71]. However, until such a solution is available, our approach can provide a feasible route for larger heterodyne HEB arrays, for example, for the proposed concept instruments HERO [32] or the far-infrared spectroscopic surveyor (FIRSS) [33].

V. CONCLUSIONS

We have successfully demonstrated three 4×2 heterodyne HEB arrays for GUSTO, which will be operated at local oscillator frequencies of 1.46, 1.9 and 4.7 THz, respectively. These arrays consist of NbN HEB mixers, where elliptical lenses and spiral antennas are applied to couple the radiation. These arrays represent, to date, the highest pixel count using the quasi-optical scheme at supra-THz frequencies. We have experimentally characterized the arrays over three key parameters, namely the mixer noise temperature ($T_{mixer}^{DSB}$), mixer conversion loss ($L_{mixer}^{DSB}$), and optimal LO power ($P_{LO}$) at the HEB. Our results demonstrate the heterodyne arrays with not only excellent sensitivity, which for example at 4.7 THz is only 6.2 times the quantum noise (hf/2k), but also good uniformity of the performance parameters. The latter is critical for efficient operation of an array within the instrument. Additionally, the measured receiver temperatures at the three frequencies, when arrays are installed in the GUSTO instrument, are also shown. GUSTO is currently in the last integration steps before being shipped to Antarctica for the launch in December 2023. Additionally, our array architecture based on quasi-optical mixers can be scaled up to a large array, e.g., 64 pixels, opening a new avenue towards large heterodyne arrays suitable for future space missions.

ACKNOWLEDGMENT

We acknowledge the technical support from Jarno Panman, Rob van der Schuur, Erik van der Meer, Henk Ode, Duc Nguyen, Marcel Dijkstra. We also thank Yuner Gan, Axel Detrain, Geert Keizer, Gabby Aitink-Kroes, Brian Jackson and Willem Jellema for helpful discussions.